\documentclass[twocolumn,
 showpacs,floats,preprintnumbers,amsmath,amssymb]{revtex4}
\usepackage{epsfig}
\usepackage{graphicx}
\usepackage{amsmath}
\usepackage{CJK}
\begin{document}
\begin{CJK*}{GBK}{song}
\title{The finite-temperature thermodynamics of a trapped unitary Fermi gas
within fractional exclusion statistics}
\author{Fang Qin\footnote{Email: qinfang@phy.ccnu.edu.cn}
and Ji-sheng Chen\footnote{Email: chenjs@iopp.ccnu.edu.cn }}
\affiliation{Physics Department and Institute of Particle Physics,
Central China Normal University, Wuhan 430079, People's Republic of
China}

\begin{abstract}

We utilize a fractional exclusion statistics of Haldane and Wu
hypothesis to study the thermodynamics
    of a unitary Fermi gas trapped in a harmonic oscillator potential at ultra-low finite temperature.
The entropy per particle
    as a function of the energy per particle and
    energy per particle versus rescaled temperature are numerically compared
    with the experimental data.
The study shows that, except the chemical potential behavior, there
exists a reasonable consistency between the
    experimental measurement and theoretical attempt for the entropy and energy per particle.
In the fractional exclusion statistics formalism, the behavior of
the isochore heat capacity for a trapped unitary Fermi gas is also
analyzed.


\end{abstract}


\maketitle
\section{Introduction}

The thermodynamics of a strongly interacting fermion system attracts
much attention, either experimentally or theoretically, in the
many-body community in recent years \cite{Giorgini2008}. For the
dilute contact interacting atomic system, the scattering length $a$
describes the mutual interaction strength between two fermions. With
the Feshbach resonance technology, it is possible to tune the
$s$-wave scattering length $a$ through changing the external
magnetic field. The scattering length $a$ can be changed from a
negative value to the other side of the resonance where $a$ becomes
positive. In the weak attraction Bardeen-Cooper-Schrieffer (BCS)
regime with $k_{F}|a|\ll1$(with the Fermi wave vector $k_{F}$), the
mean field theory can give a reasonable description. With the
increase of the interaction strength, the composite bosons can be
formed in the fermion system, where the Bose-Einstein condensation
(BEC) can occur \cite{Giorgini2008,Ho2004}.

In the regime from negative scattering length to positive value, the
two-body interaction strength can be singular with the existence of
a zero-energy bound state. The limit $k_{F}|a|\rightarrow\infty$ is
called BCS-BEC crossover, which is also called the unitary regime
\cite{Giorgini2008}. The corresponding gas in this limit is called
unitary Fermi gas \cite{Ho2004}. At unitarity, the gas both has some
properties of fermions and some of bosons, that is to say, the
behavior of this gas is between that of fermions and bosons. The
fermion system will show a universal thermodynamic behavior near the
resonance point where the scattering length $a$ diverges
\cite{Giorgini2008,Ho2004}.

Initially, physicists concentrated on the energy density per
particle of the unitary Fermi gas at zero temperature theoretically.
For a homogeneous gas, the ground-state energy per particle is given
by $E/N=\frac{3}{5}\xi E_{F}$ with the ideal Fermi energy $E_{F}$
and the universal coefficient $\xi$ \cite{Ho2004}. It is
$E/N=\frac{3}{4}\sqrt{\xi}E_{F}$ for a trapped unitary gas, where
$E_{F}$ is the Fermi energy of the trapped noninteracting Fermi gas
\cite{Thomas2005}. Determining the dimensionless constant $\xi$ has
been an important physical subject in the past few years.

Furthermore, exploring the thermodynamics of a unitary Fermi gas at
finite temperature in theory to explain the experimental results is
a more challenging task. A diagrammatic determinant Monte Carlo
method for the negative-$U$ Hubbard model was used to calculate the
finite-temperature thermodynamics of a homogeneous unitary gas
\cite{DDMC}. Through the quantum Monte Carlo simulation, the
finite-temperature thermodynamics of a homogeneous unitary gas can
be given by Refs.\cite{QMC1,QMC2}. A self-consistent theory based on
the combined Luttinger-Ward-De Dominicis-Martin variational
formalism was also used to calculate it for a homogeneous unitary
gas \cite{Haussmann1}.

For the finite-temperature thermodynamics of a trapped unitary Fermi
gas, the entropy and energy of the trapped unitary Fermi system have
been experimentally measured in Refs.\cite{Kinast,Luo2007,Luo2009}.
The quantum Monte Carlo simulation can also be used to calculate the
finite-temperature thermodynamics of a trapped unitary gas
\cite{QMC3}. Various theoretical attempts have been established to
give the calculations for a trapped unitary Fermi gas. The pseudogap
theory is one of them \cite{Chen1,Chen2}. A mean-field method by
solving the Bogoliubov-de Gennes equations with an efficient and
accurate method was used in Ref.\cite{Hu1}. A $T$-matrix calculation
with a modified Nozi$\grave{e}$res and Schmitt-Rink approximation
was adopted to explore it \cite{Hu2}. The combined Luttinger-Ward-De
Dominicis-Martin variational formalism was also proved to be a
valuable approach for it \cite{Haussmann2}. We note that there are
still differences among in these attempts on the finite-temperature
thermodynamic properties of a unitary Fermi gas.

Due to the scale invariance at unitarity, the thermodynamic
properties of the universal strongly interacting unitary fermions
are related to those of the non-interacting ideal fermions. This
means that the dynamical details will not appear in the final
thermodynamic analytical expressions. From the microscopic point of
view, the unitary fermions system is in between the fermionic and
composite bosonic phases. To characterize this crossover or
intermediate unitary fermionic thermodynamics, Haldane-Wu fractional
exclusion statistics was used to discuss the finite-temperature
unitary Fermi gas thermodynamics \cite{Bhaduri1,Bhaduri2}. In
physics, the thermodynamic behavior of the fractional exclusion
statistics is between those of Bose-Einstein and Fermi-Dirac
statistics \cite{Haldane1991,Wu1994}, which is quite similar to the
thermodynamics of real fermions at unitarity. As a hypothesis, the
strongly interacting unitary gas is modeled by the non-interacting
anyons. The priority is that the finite-temperature thermodynamic
properties can be investigated analytically \cite{Sevincli}. The
quintessence hidden in this attempt is that the anyonic statistical
parameter $g$ characterizes the strongly interacting universal
properties.

Following the fractional exclusion statistics formalism, the aim of
this work is to give a detailed discussion on the trapped gas
thermodynamics. By generalizing the discussion for the homogeneous
unitary system \cite{Qin}, we want to give the expressions and the
corresponding numerical results of entropy, physical chemical
potential and isochore heat capacity of a trapped unitary system.
The concrete comparisons with experimental measurements are made.

The natural units $k_{B}=\hbar=1$ are used throughout the paper.

\section{Finite-temperature thermodynamics within fractional
exclusion statistics}

\subsection{Framework of fractional exclusion statistics}

In this subsection, we will briefly review the formalism of the
fractional exclusion statistics. By generalizing the simple formula
with the fractional exclusion statistics, one can get the
microscopic quantum states $W$ of $N$ identical particles occupying
a group of $G$ states \cite{Wu1994,Qin} \begin{eqnarray}\label{W1}
W=\prod_{i}\frac{[G_{i}+(N_{i}-1)(1-g)]!}{N_{i}![G_{i}-gN_{i}-(1-g)]!},
\end{eqnarray} where $g$ is a statistical parameter, which denotes the number
of states that one particle can occupy. For bosons, $g=0$ and $g=1$
for fermions.

If $G_{i}$ and $N_{i}$ are very large, one has the approximate
expression of the logarithm of $W$ \begin{eqnarray}\label{W2}
\ln{W}&&\simeq\sum_{i}\left
[(G_{i}+(1-g)N_{i})\ln{(G_{i}+(1-g)N_{i})}
\right. \nonumber\\
&& \left. ~~-(G_{i}-gN_{i})\ln{(G_{i}-gN_{i})}-N_{i}\ln{N_{i}}\right
],
\end{eqnarray} through the Stirling formula $\ln{N!}=N(\ln{N}-1)$.

The variational formulation of $\ln{W}$ is
\begin{eqnarray}\label{W3}
\delta\ln{W}&&=\sum_{i}\left [(1-g)\ln{(G_{i}+(1-g)N_{i})}
\right. \nonumber\\
&& \left. ~~+g\ln{(G_{i}-gN_{i})}-\ln{N_{i}}\right ]\delta N_{i}.
\end{eqnarray}

However, $\delta N_{i}$ is not arbitrary, it must satisfy the
conditions below:
\begin{eqnarray} \delta N&=&\sum_{i}\delta N_{i}=0, \nonumber\\
              \delta E&=&\sum_{i}\epsilon_{i}\delta N_{i}=0.
\end{eqnarray}

Setting the two Lagrange multipliers $\alpha=-\mu/T$ and
$\beta=1/T$, one can have \begin{eqnarray}\label{W4}
&&\delta\ln{W}-\alpha\delta N-\beta\delta
E \nonumber\\
=&&\sum_{i}\left [(1-g)\ln{(G_{i}+(1-g)N_{i})}
\right. \nonumber\\
&& \left.
+g\ln{(G_{i}-gN_{i})}-\ln{N_{i}}+(\mu-\epsilon_{i})/T\right ]\delta N_{i}\nonumber\\
=&&\sum_{i}\left [(1-g)\ln{\left (1+\frac{G_{i}}{N_{i}}-g\right )}
\right. \nonumber\\
&& \left. +g\ln{\left (\frac{G_{i}}{N_{i}}-g\right )}+\left
(\frac{\mu-\epsilon_{i}}{T}\right )\right ]\delta N_{i},
\end{eqnarray} where $\mu$ is the chemical potential, $T$ is the
system temperature and $\epsilon_{i}$ is the single-particle energy
for the state of species $i$.

Defining the average occupation number $\bar{N_{i}}\equiv
N_{i}/G_{i}$, through the Lagrange multiplier method
$\delta\ln{W}-\alpha\delta N-\beta\delta E=0$, the most probable
distribution of $\bar{N_{i}}$ can be derived from Eq.(\ref{W4})
\begin{eqnarray}\label{f} \bar{N_{i}}=\frac{1}{\omega_{i}+g}, \end{eqnarray}
where $\omega_{i}$ obeys the relation \begin{eqnarray}\label{o}
\mu-\epsilon_{i}=-T\left
[(1-g)\ln{(1+\omega_{i})}+g\ln{\omega_{i}}\right ].
\end{eqnarray}

In Eq.(\ref{o}),  we define $\omega_{0}$ as the value of
$\omega_{i}$ at $\epsilon_{i}=0$. Consequently, one has
\begin{eqnarray}\label{mu0}
\mu=-T[(1-g)\ln{(1+\omega_{0})}+g\ln{\omega_{0}}]. \end{eqnarray}

At zero temperature, there is
\begin{eqnarray}\label{f0} \bar{N_{i}}&=&0~,~~for~\epsilon_{i}>\mu; \nonumber\\
              \bar{N_{i}}&=&\frac{1}{g}~,~~for~\epsilon_{i}<\mu.
\end{eqnarray}

Furthermore, by inserting Eq.(\ref{f}) into Eq.(\ref{W2}), one
obtains the expression for entropy as \begin{eqnarray}\label{S}
S=\ln{W}=\sum_{i}\frac{G_{i}}{\omega_{i}+g}[(\omega_{i}+1)\ln{(\omega_{i}+1)}-\omega_{i}\ln{\omega_{i}}].
\end{eqnarray}

\subsection{Thermodynamics of a trapped unitary Fermi gas}

The finite-temperature thermodynamics can be derived for the unitary
Fermi gas trapped in a harmonic oscillator $m\varpi^{2}r^{2}/2$,
where the oscillator parameter $\varpi$ is defined as
$\varpi=(\omega_{x}\omega_{y}\omega_{z})^{1/3}$ and $r$ denotes the
position of particles. The corresponding density of states is
$D(\epsilon)=\epsilon^{2}/\varpi^{3}$. At zero temperature, the
particle number and system energy are
\begin{eqnarray}\label{N0}
N=\frac{1}{g}\int_{0}^{\widetilde{E}_{F}}{D(\epsilon)d\epsilon}=\frac{E_{F}^{3}}{3\varpi^{3}},
\end{eqnarray} \begin{eqnarray}\label{E0}
E=\frac{1}{g}\int_{0}^{\widetilde{E}_{F}}{\epsilon
D(\epsilon)d\epsilon}=\frac{g^{1/3}E_{F}^{4}}{4\varpi^{3}},
\end{eqnarray} where $\widetilde{E}_{F}$ obeys the relation
$\widetilde{E}_{F}=g^{{1}/{3}}E_{F}$ with the ideal Fermi energy
$E_{F}=(3N)^{1/3}\varpi$ in a harmonic oscillator.

Comparing Eq.(\ref{N0}) with Eq.(\ref{E0}) to eliminate $\varpi$, it
is shown that
\begin{eqnarray}\label{xi} \frac{E}{NE_{F}}=\frac{3}{4}g^{1/3}.
\end{eqnarray}

If the statistical parameter $g$ in the fractional exclusion
statistics is fixed through the zero-temperature ground-state
energy, the finite-temperature thermodynamic quantities for a
trapped unitary Fermi gas can be calculated. For the trapped system,
the ground-state energy is related with the universal coefficient
according to $E/(\frac{3}{4}NE_{F})=\sqrt{\xi}$ \cite{Thomas2005}.
From Eq.(\ref{xi}), it is found that
$E/(\frac{3}{4}NE_{F})=g^{1/3}$. So the statistical parameter could
be calculated as $g=\xi^{3/2}=\frac{8}{27}$, with $\xi=\frac{4}{9}$
given by the developed quasi-linear approximation method
\cite{Chenjs1,Chenjs2,Chenjs3,chen2007,chen20072,chen2009}.

For the finite-temperature trapped unitary Fermi system, the
particle number and energy can be represented by turning the sum of
quantum state into integral and changing the variable from
$d\epsilon$ to $d\omega$
\begin{eqnarray}\label{N1}
N&&=\sum_{i}G_{i}\bar{N_{i}}=\int_{0}^{\infty}{\frac{D(\epsilon)d\epsilon}{\omega+g}}
=\left (\frac{T}{\varpi}\right )^{3}h_{2}(\omega_{0}),\\
\label{E1}
E&&=\sum_{i}G_{i}\bar{N_{i}}\epsilon_{i}=\int_{0}^{\infty}{\frac{\epsilon
D(\epsilon)d\epsilon}{\omega+g}}=\left (\frac{T}{\varpi}\right
)^{3}Th_{3}(\omega_{0}),
\end{eqnarray} where \begin{eqnarray}
h_{n}(\omega_{0})&&=\int_{\omega_{0}}^{\infty}{\frac{d\omega}{\omega(1+\omega)}\left
[\ln{\left (\frac{\omega}{\omega_{0}}\right )^{g}\left
(\frac{1+\omega}{1+\omega_{0}}\right )^{1-g}}\right ]^{n}}.\nonumber
\end{eqnarray}

By replacing Eq.(\ref{N0}) into Eq.(\ref{N1}) and Eq.(\ref{E1}), one
gets
\begin{eqnarray}\label{N2}
&&3\left (\frac{T}{T_{F}}\right )^{3}h_{2}(\omega_{0})=1,\\
\label{E2} &&\frac{E}{NE_{F}}=3\left (\frac{T}{T_{F}}\right
)^{4}h_{3}(\omega_{0}),
\end{eqnarray} with the Fermi characteristic temperature $T_{F}$.

One can turn the sum of Eq.(\ref{S}) to an integral. With
Eq.(\ref{N2}), one can get the explicit integral expression of
entropy per particle
\begin{eqnarray}\label{S1}
\frac{S}{N}&&=3\left (\frac{T}{T_{F}}\right )^{3}\int_{\omega_{0}}^{\infty}\left [\left (\frac{\ln{(\omega+1)}}{\omega}-\frac{\ln{\omega}}{\omega+1}\right )\right. \nonumber\\
&& \left. ~~\times\left (\ln{\left (\frac{\omega}{\omega_{0}}\right
)^{g}\left (\frac{1+\omega}{1+\omega_{0}}\right )^{1-g}}\right
)^{2}\right ]d\omega.
\end{eqnarray}

In order to derivate the expression of the isochore heat capacity,
let us further discuss the particle number. From Eq.(\ref{N1}), the
partial derivative of the particle number $N$ to temperature $T$ for
fixed $N$ and $\varpi$ is given by
\begin{eqnarray}\label{*} \left (\frac{\partial N}{\partial
T}\right )_{N,\varpi}&&=\frac{3T^{2}}{\varpi^{3}}h_{2}(\omega_{0})+\left (\frac{T}{\varpi}\right )^{3}\left [\left (\frac{\partial h_{2}}{\partial T}\right )_{\mu}\right. \nonumber\\
&& \left. ~~+\left (\frac{\partial h_{2}}{\partial \mu}\right
)_{T}\left (\frac{\partial \mu}{\partial T}\right )_{N,\varpi}\right
].
\end{eqnarray} From Eq.(\ref{*}), one can get \begin{eqnarray}\label{muN}
\left (\frac{\partial \mu}{\partial T}\right
)_{N,\varpi}=\frac{\mu}{T}-\frac{3h_{2}(\omega_{0})}{2h_{1}(\omega_{0})}.
\end{eqnarray}

Furthermore, with Eq.(\ref{E1}) and the thermodynamic relation
between the isochore heat capacity $C_{V}$ and internal energy $E$,
one has
\begin{eqnarray}\label{CV}
C_{V}&&\nonumber=\left (\frac{\partial E}{\partial T}\right )_{N,\varpi}\\
\nonumber&&=4\left (\frac{T}{\varpi}\right )^{3}h_{3}(\omega_{0})
+\left (\frac{T}{\varpi}\right )^{3}T\left [\left (\frac{\partial
h_{3}}{\partial T}\right )_{\mu}\right. \nonumber\\
&& \left.+\left (\frac{\partial h_{3}}{\partial \mu}\right
)_{T}\left (\frac{\partial \mu}{\partial T}\right )_{N,\varpi}\right
].
\end{eqnarray} Substituting Eq.(\ref{N1}) and Eq.(\ref{muN}) into Eq.(\ref{CV}), one
obtains the isochore heat capacity per particle
\begin{eqnarray}\label{CV1}
\frac{C_{V}}{N}=\frac{4h_{3}(\omega_{0})}{h_{2}(\omega_{0})}-\frac{9h_{2}(\omega_{0})}{2h_{1}(\omega_{0})}.
\end{eqnarray}

\section{Discussions}

Based on the above analytical expressions, we will give the
numerical results.

\begin{figure}[htb]
  \centering
   \includegraphics[width =0.45\textwidth]{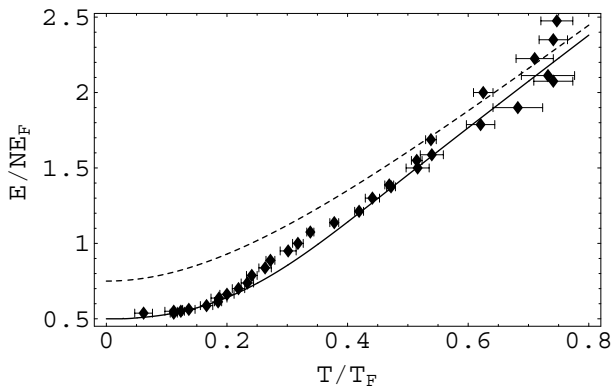}
   \caption{\small
    The energy per particle versus the rescaled
temperature. The solid curve denotes that for the trapped unitary
Fermi gas, and the dashed one is that for the trapped ideal Fermi
gas. The dots with error bars are the experimental data of
Refs.\cite{Kinast}. }\label{fig1}
\end{figure}

The energy per particle versus the rescaled temperature can be
calculated from Eq.(\ref{N2}) and Eq.(\ref{E2}). As indicated by the
Fig.\ref{fig1}, the energy for the trapped unitary Fermi gas with
the statistical parameter $g=\frac{8}{27}$ versus the rescaled
temperature is consistent with the experimental data \cite{Kinast}.

\begin{figure}[htb]
  \centering
   \includegraphics[width =0.45\textwidth]{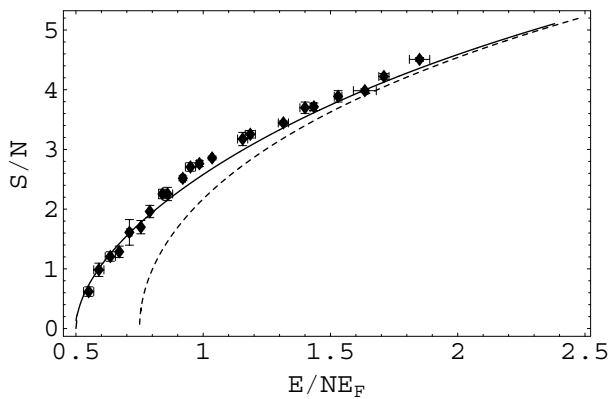}
   \caption{\small
    The entropy per particle plotted as
    a function of the rescaled energy
per particle. The line styles are similar to Fig.\ref{fig1}. The
dots with error bars are the experimental data of
Refs.\cite{Luo2007,Luo2009}. }\label{fig2}
\end{figure}

Through Eq.(\ref{E2}) and Eq.(\ref{S1}), the relation between the
entropy per particle and the energy per particle is given explicitly
in Fig.\ref{fig2}. The entropy increases with increasing energy, and
in the Boltzmann regime, the curve for a trapped unitary Fermi gas
gets closer to and almost overlaps with that of the trapped ideal
Fermi gas. It is found that the theoretical result is reasonably
consistent with that of the experiment.

The agreements in Fig.\ref{fig1} and Fig.\ref{fig2} between the
theoretical curves and the experimental data show that the
statistical parameter $g$ can be capable of describing the strong
interaction of unitary Fermi gas at extremely low temperature.

\begin{figure}[htb]
  \centering
   \includegraphics[width =0.45\textwidth]{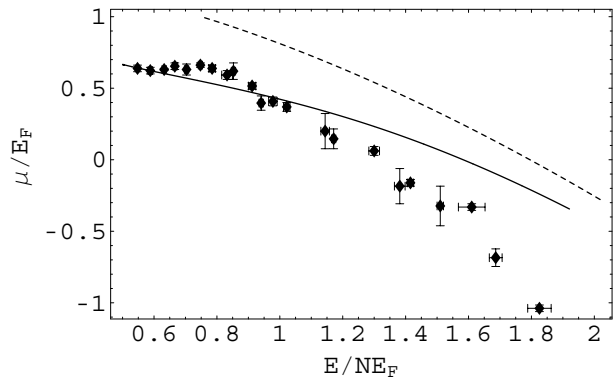}
   \caption{\small
    Similar to Fig.\ref{fig1} for the variation
of the chemical potential as a function of the rescaled energy per
particle. The dots with error bars are the experimental data of
Ref.\cite{Luo2009}. }\label{fig3}
\end{figure}

\begin{figure}[htb]
  \centering
   \includegraphics[width =0.45\textwidth]{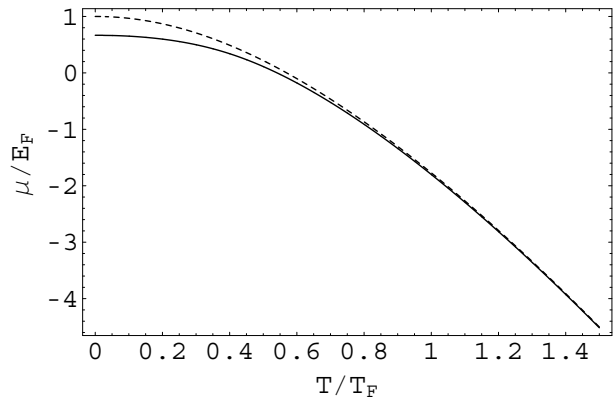}
   \caption{\small
    The relationship between chemical potential and the recaled
temperature. The line styles are the same as in Fig.\ref{fig1}.
}\label{fig4}
\end{figure}

With Eqs.(\ref{mu0}), (\ref{N2}) and (\ref{E2}), we have also shown
the plot of the chemical potential varying with the rescaled energy
per particle in Fig.\ref{fig3}. It is a monotonically decreasing
function of the energy per particle. In the Boltzmann regime, the
curve for a trapped unitary Fermi gas gets closer to that of the
trapped ideal one. In the extremely low-temperature regime
($E/(NE_{F})<1.2$), the departure between the theoretical result and
the experimental data is not obvious. However, the chemical
potential differs explicitly from the experimental result for
$E/(NE_{F})>1.2$. As the temperature increases, the experimental
data also diverge from the curve of the trapped ideal gas. We can
also plot the chemical potential versus the rescaled temperature in
Fig.\ref{fig4}.

\begin{figure}[htb]
  \centering
   \includegraphics[width =0.45\textwidth]{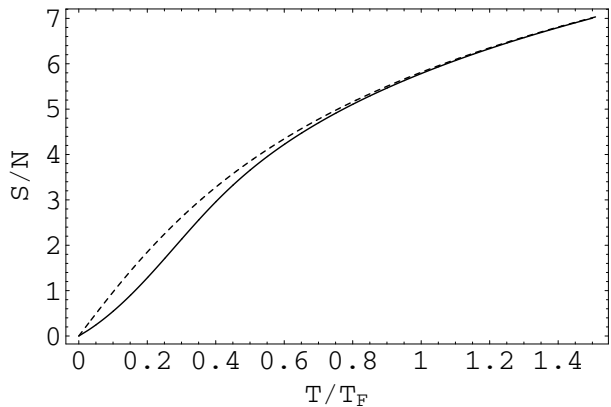}
   \caption{\small
    Same as in Fig.\ref{fig4} for entropy per particle plotted as
    a function of the recaled temperature. }\label{fig5}
\end{figure}

\begin{figure}[htb]
  \centering
   \includegraphics[width =0.45\textwidth]{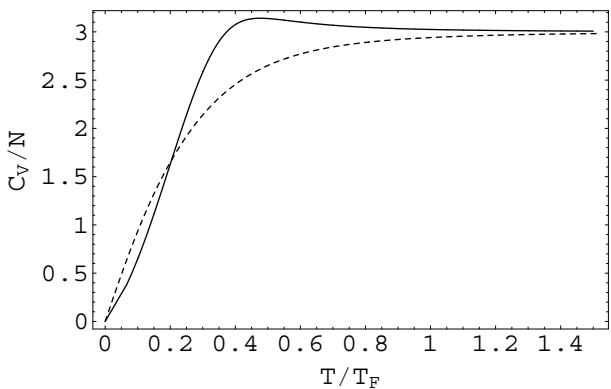}
   \caption{\small
   Plot of the isochore heat capacity per particle as a function of the
recaled temperature. The line styles are the same as those shown in
        Fig.\ref{fig5}. }\label{fig6}
\end{figure}

From Eqs.(\ref{N2}) and (\ref{S1}), the entropy per particle as a
function of the rescaled temperature is indicated in Fig.\ref{fig5}.
In Fig.\ref{fig6}, the isochore heat capacity per particle related
with the rescaled temperature is presented. The numerical results
are obtained by solving the coupled Eqs.(\ref{N2}) and (\ref{CV1}).

Near the Boltzmann regime, the energy, chemical potential, entropy
and isochore heat capacity of a trapped unitary Fermi gas are
getting closer to and almost overlap with those of the trapped ideal
Fermi gas. In the low-temperature strong degenerate regime, the
thermodynamic quantities of the unitary gas are lower than the ones
of the ideal gas at the same temperature. The energy, chemical
potential and entropy of a unitary gas are always lower than the
ideal ones at the same temperature. However, as shown in
Fig.\ref{fig6}, the isochore heat capacity per particle given by the
fractional exclusion statistics is not a naive monotonously
increasing function with the increase of the scaled temperature.
This behavior is different from that of the trapped ideal Fermi gas.

\section{Conclusion}

The finite-temperature thermodynamic quantities of a trapped unitary
Fermi gas have been discussed in terms of the approved fractional
exclusion statistics framework. The study shows that the
thermodynamic quantities of a trapped unitary Fermi gas will overlap
with those of the trapped ideal Fermi gas in the Boltzmann regime.
The thermodynamic quantities given by this formalism are lower than
the trapped ideal ones at the same temperature except for the
isochore heat capacity.

The energy and entropy per particle manifest the consistency with
the low-temperature experimental measurement. However, the detailed
numerical study of the chemical potential shows that there is an
explicit difference between the result given in terms of fractional
exclusion statistics and experimental data in the weakly degenerate
regime. For the three-dimensional trapped unitary Fermi gas, the
high-temperature weakly degenerate behavior of chemical potential
given by a simple fractional exclusion statistics hypothesis needs
clarifying further.

\acknowledgments{ Supported in part by the National Natural Science
Foundation of China under Grant No. 10675052 and 10875050 and MOE of
China under projects No.IRT0624. }

\end{CJK*}
\end{document}